\documentclass[a4paper]{article}

\usepackage{INTERSPEECH2022}
\usepackage{xcolor}
\usepackage{makecell}
\usepackage{multirow}
\usepackage{comment}
\usepackage{array}
\usepackage{changepage}
\usepackage{caption}
\newcolumntype{?}{!{\vrule width .1pt}}

\title{Simple and Effective Multi-sentence TTS \\ with Expressive and Coherent Prosody}
\name{
    Peter Makarov$^{*\dagger}$\thanks{\scriptsize{$^*$ Equal contribution.}}\thanks{\scriptsize{$^\dagger$ Work done while at Amazon.}},
    Ammar Abbas$^*$,
    Mateusz Lajszczak,
    Arnaud Joly,
    Sri Karlapati, \\
    Alexis Moinet,
    Thomas Drugman,
    Penny Karanasou
}
\address{Alexa AI, Amazon}
\email{syeabbs@amazon.com}

\begin{document}

\maketitle
\begin{abstract}

Generating expressive and contextually appropriate prosody remains a challenge for modern text-to-speech (TTS) systems. This is particularly evident for long, multi-sentence inputs.
In this paper, we examine simple extensions to a Transformer-based FastSpeech-like system, with the goal of improving prosody for multi-sentence TTS.
We find that long context, powerful text features, and training on multi-speaker data all improve prosody. More interestingly, they result in synergies. Long context disambiguates prosody, improves coherence, and plays to the strengths of Transformers. Fine-tuning word-level features from a powerful language model, such as BERT, appears to benefit from more training data, readily available in a multi-speaker setting. 
We look into objective metrics on pausing and pacing and perform thorough subjective evaluations for speech naturalness. 
Our main system, which incorporates all the extensions, achieves consistently strong results, including statistically significant improvements in speech naturalness over all its competitors.
\end{abstract}


\section{Introduction}

Recent advances in neural TTS \cite{Wangetal2017,Shenetal2018,Renetal2019,Renetal2021} have unlocked a range of applications in which human-like and coherent prosody is crucial for customer experience. This is particularly the case for applications involving complex multi-sentence input, such as conversational agents or systems reading out news or Wikipedia articles. 
Most research in TTS has focused on training and evaluating models on single, isolated sentences. While the advantage of this approach is that it applies to any speech dataset, it might not work as well in domains with multi-sentence input.
For example, \cite{Clarketal2019} demonstrate that the perceptual evaluation score of a paragraph cannot be reliably predicted from the scores of its sentences evaluated in isolation. Prosodic coherence and contextual appropriateness are thus highly important and should be modeled accordingly.

In this paper, we focus on TTS for domains with complex multi-sentence input, which require expressive and coherent prosody, e.g.\ long-form reading, dialogues, question-answering prompts, etc. We investigate what improvements to speech quality could be achieved through the following simple extensions to a single-sentence baseline model:

\begin{enumerate}
    \item \textbf{long context}: training and synthesis on utterances containing multiple adjacent coherent input sentences
    
    \item \textbf{contextual word embeddings}: conditioning on contextual word embeddings \cite{Devlinetal2019}, which enable context disambiguation at a more coarse-grained level than frames or phonemes,
    
    \item \textbf{multi-speaker modeling}, which facilitates transfer learning and helps avoid overfitting due to larger combined training dataset sizes.
\end{enumerate}

Naturally, these extensions can be applied simultaneously to the same model.
We base our work on a \textbf{Transformer}-based FastSpeech-like TTS system \cite{Renetal2019,Renetal2021}, which achieves high speech quality, efficiency, and is generally expected to perform best on longer inputs.
We examine how these extensions---both individually and in various combinations---impact overall prosody and its coherence on multi-sentence input. We look into objective metrics related to pausing and pacing, and perform thorough subjective evaluations for speech naturalness.

We summarize our contributions and findings below.
\begin{enumerate}
    \item We examine three simple extensions to a state-of-the-art TTS system to achieve more expressive, contextually appropriate, and coherent prosody on multi-sentence input: long context, contextual word embeddings, and multi-speaker modeling. To the best of our knowledge, we are the first to consider them in this setting together.

    \item We see gains from all three extensions. Contextual word embeddings and long context both lead to the largest improvements in naturalness and metric scores for pausing and pacing. On the other hand, synthesis on multi-sentence input using a model not trained with long context does not work well.
    
    \item We observe synergy effects. In particular, the baseline model extended in all three ways at once
    outperforms other system combinations in naturalness MUSHRA~\cite{series2014method} evaluations by a margin. For example, it achieves a statistically significant gap reduction%
    \footnote{%
    \scriptsize{The gap reduction in mean MUSHRA scores between systems~1 and 2 given reference is computed as $100*(\textrm{system1} - \textrm{system2}) / (\textrm{reference} - \textrm{system2})$.}}
    of 59\% over multi-speaker Transformer-BERT.

\end{enumerate}

\section{Related work}

Although most research effort in TTS has focused on models operating on isolated sentences, there has been growing interest in improving synthesis for longer, multi-sentence inputs.  Most innovations focus on better modeling of surrounding textual content. This is achieved with generic paragraph-based features \cite{Prahalladetal2007,Peiroetal2018}, hand-crafted discourse relation labels \cite{Aubinetal2019,Huetal2016}, and---more recently---textual embeddings of neighboring sentences \cite{Xuetal2021,Wangetal2020}. \cite{Oplustiletal2020} study the impact of previous acoustic context. In this paper, we explore a most direct approach to long context, which involves both text and mel-spectrogram context conditioning.

Transformers \cite{Vaswanietal2017} are particularly well suited for modeling long-distance dependencies in the data as they do not suffer from the recency bias, unlike earlier models. This is advantageous for prosody modeling, particularly with multi-sentence inputs. %
\cite{Renetal2019,Renetal2021} apply non-autoregressive Transformers to TTS, and we build on their work.

There has been a lot of interest in multi-speaker TTS recently~\cite{Gibianskyetal2017,Jiaetal2018,Chenetal2019,Chenetal2020,Cooperetal2020}. Increased scalability and improvements through transfer learning are the main advantages of such models. Unlike previous work, we examine to which extent transfer learning with multi-speaker modeling can help improve prosody and coherence of multi-sentence speech.

\cite{Hayashietal2019,Fangetal2019,Prateeketal2019,Kenteretal2020,Xiaoetal2020,Jiaetal2021} apply contextual word embeddings from large pre-trained language models in the TTS backend. Most closely related to our work is \cite{Kenteretal2020}, who show that fine-tuned BERT word embeddings improve the perception of synthesized speech. Building on this work, we further investigate these representations in a multi-speaker setting and in the presence of longer context.

\section{Methods}

\begin{figure}
    \captionsetup{font=footnotesize}
    \centering
    \includegraphics[scale=.6]{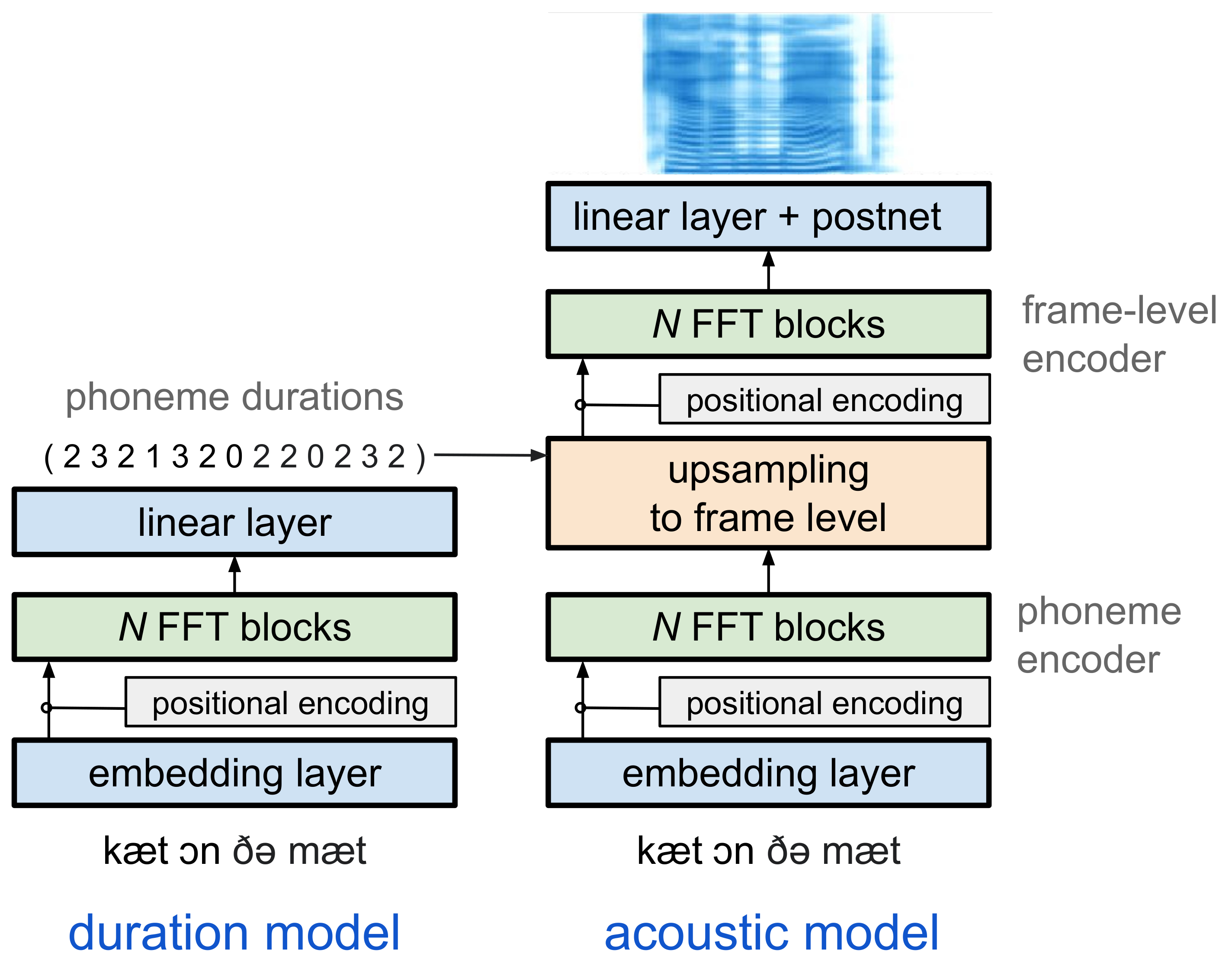}
    \caption{FastSpeech-like Transformer baseline.}
    \label{fig:baseline}
\end{figure}

\subsection{FastSpeech-like Transformer baseline}
\label{subsec:baseline}
We take a Transformer-based FastSpeech-like system \cite{Renetal2019} as our baseline. The system is comprised of an acoustic model and a duration model (Figure~\ref{fig:baseline}). Both models use Feed-Forward Transformer (FFT) blocks as feature encoders.
The duration model predicts phoneme durations (in frames) and is a regressor on top of the phoneme FFT encoder. The acoustic model predicts mel-spectrogram frames and consists of the phoneme FFT encoder and frame-level feature FFT encoder. The output embeddings of the phoneme FFT encoder are upsampled to the frame level using the phoneme durations predicted by the duration model and then fed into the frame-level feature FFT encoder. The acoustic and duration models use separate phoneme encoders. The architecture of the duration model and the fact that we use force-aligned phoneme durations as training targets are the two main differences from FastSpeech.

\subsection{Extensions}

We examine three extensions to this baseline with the goal of improving expressiveness and prosodic coherence on multi-sentence inputs.

\subsubsection{Multi-speaker modeling}
\label{subsec:multi}
Multi-speaker TTS systems are commonly used as a data reduction technique, with the expectation of obtaining strong positive transfer from high-resource to low-resource speakers \cite{Gibianskyetal2017,Chenetal2020}.
In the context of large language models (LM) for TTS (see \S\ref{subsec:bert} next), the added advantage of multi-speaker modeling is that it reduces the risk of overfitting the LLM during fine-tuning simply due to the sheer amount of training data. To turn the baseline into a multi-speaker model, we concatenate pre-trained speaker embeddings to phoneme encodings. The resulting feature vectors are passed through a ReLU layer (to match the FFT block input dimension) and upsampled to the frame level. The speaker embeddings are obtained from a speaker verification model~\cite{Wanetal2018}.

\subsubsection{BERT word embeddings}
\label{subsec:bert}

\begin{figure}
    \captionsetup{font=footnotesize}
    \centering
    \includegraphics[scale=0.43]{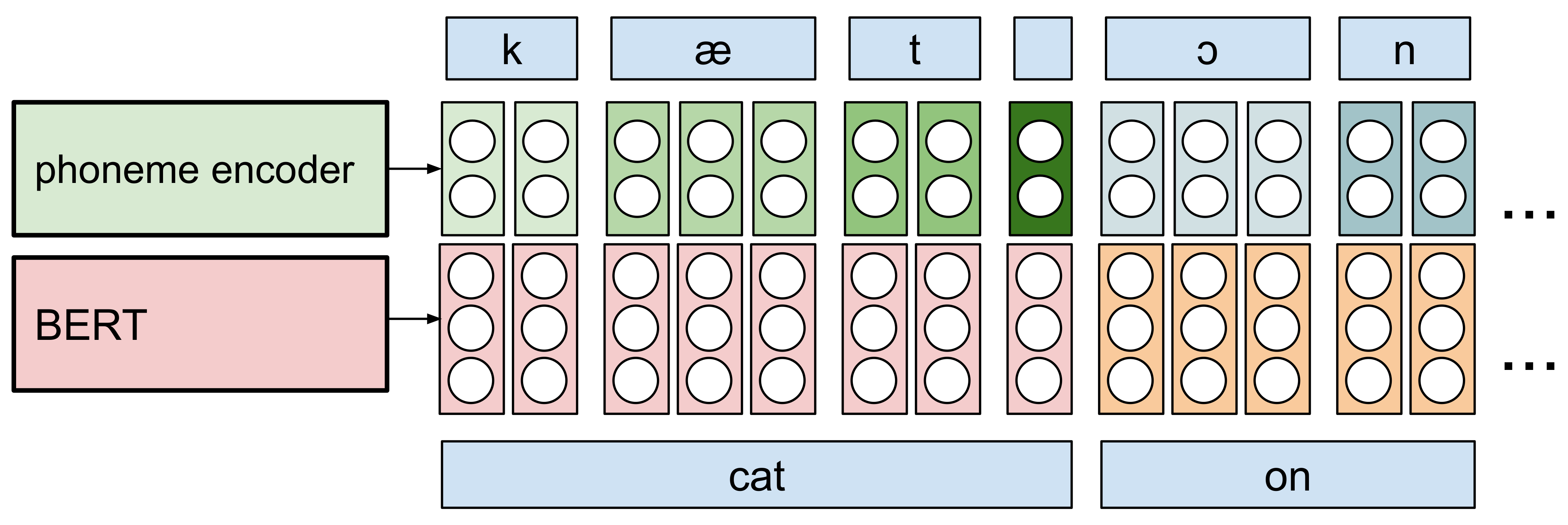}
    \caption{Alignment and upsampling of phoneme encodings (top vectors, shares of {\color{teal}green}) and BERT word embeddings (bottom vectors, shades of {\color{red}red}) in Transformer-BERT.}
    \label{fig:bert}
\end{figure}

Since prosody closely tracks syntax \cite{Bennett&Elfner2019,Kohnetal2018}, linguistic features (especially syntactic information) are widely believed to be beneficial for TTS. In line with previous work \cite{Kenteretal2020}, we extend both the duration and acoustic models with contextual word embeddings from a large pre-trained Transformer LM. Concretely, we concatenate the word embedding to the encodings of the phonemes making up this word (Figure~\ref{fig:bert}). As with speaker embeddings, the resulting feature vectors pass through a ReLU layer before being upsampled to the frame-level. The word embedding is computed by taking the output embedding corresponding to the first sub-token of that word. We use the cased version of BERT-base \cite{Devlinetal2019} and fine-tune it on the end task (phoneme duration or mel-spectrogram frames prediction) as we train the rest of the model. We refer to this baseline extension as Transformer-BERT.

\subsubsection{Long context}

TTS systems are generally built to work with single-sentence inputs. Some advantages of this approach are that it applies to any dataset and that synthesis is trivially parallelizable across sentences. Yet, in case data consists of coherent multiple sentences (e.g.\ dialogues, long-form reading, question answering prompts), it may be advantageous to take longer context into account. 
Long context disambiguates prosody, which is particularly evident for utterances that would be shorter otherwise. It leads to a simpler learning problem, which is beneficial for TTS models without latent variables, and narrows the space of possible realizations at synthesis time.

Here, we take a simple approach to incorporating long context for domains with multi-sentence input. We propose training and synthesizing on utterances spanning multiple adjacent sentences. To this end, we concatenate sentences that come after each other in the original data. We highlight that this approach is viable for non-attention systems such as the baseline. Attention-based models are known to suffer from instabilities arising from attention such as mumbling, skipping, or repetitions in the audio \cite{Renetal2019,Yuetal2019}, which can deteriorate with input length.

\begin{figure}
    \captionsetup{font=footnotesize}
    \centering
    \includegraphics[scale=.53]{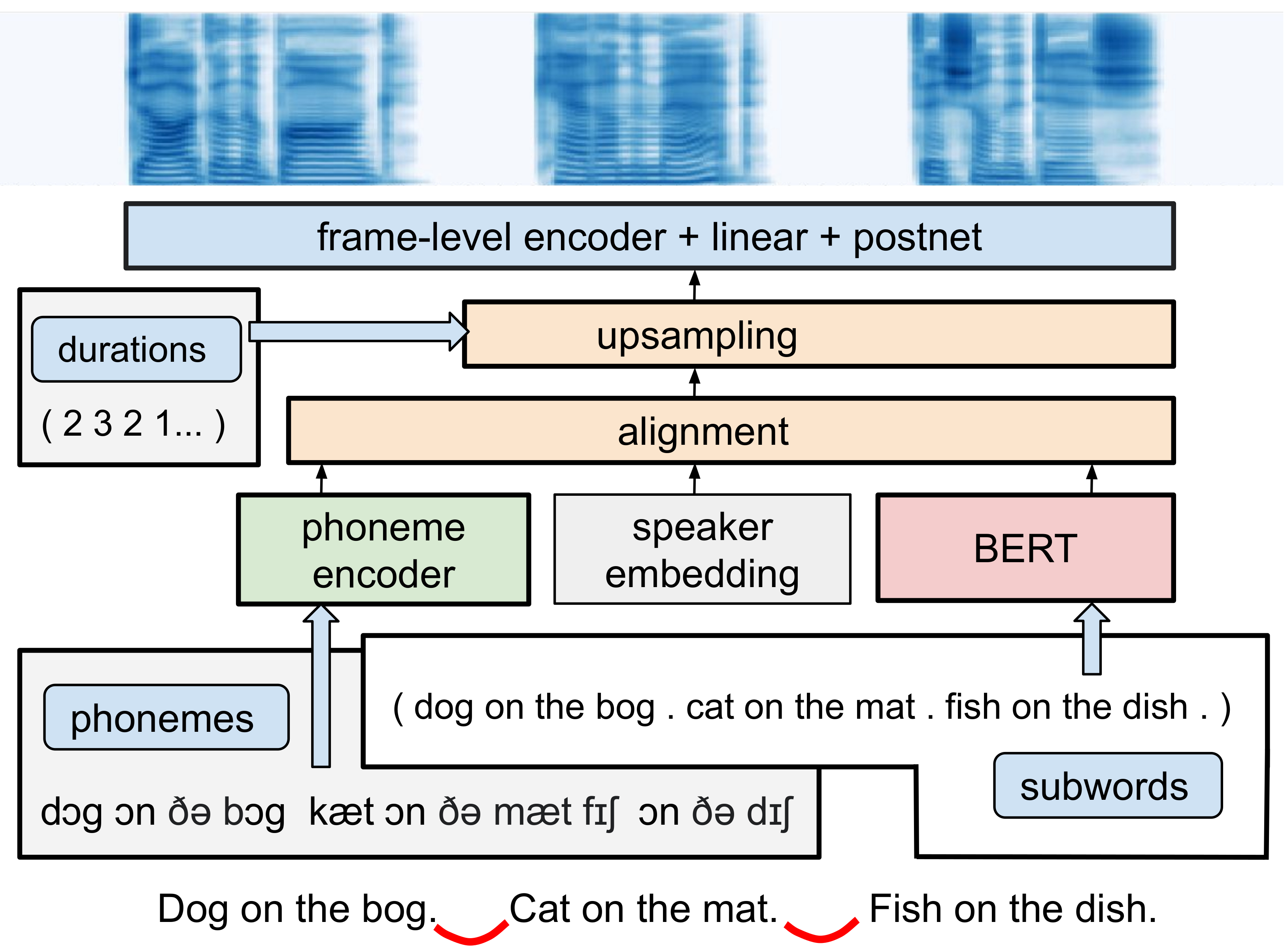}
    \caption{Multi-speaker long-context Transformer-BERT.}
    \label{fig:context}
\end{figure}

We chunk the training data into consecutive chunks of concatenated sentences.
A chunk is always composed of complete sentences, and there is a limit on the maximum length of the chunk.
We experimented with multiple maximum lengths and found that approximately 24 seconds worked well with our data.
This takes into account training efficiency and model convergence as longer utterances put a significant strain on batch size, particularly when combined with fine-tuned word embeddings. We decided against chunks of fixed length, which would lead to incomplete sentences, since this would potentially increase ambiguity in prosody. Figure~\ref{fig:context} visualizes the incorporation of long context with the rest of the components.

\begin{table}[!b]
    \captionsetup{font=footnotesize}
    \centering
    \resizebox{.47\textwidth}{!}{
    \begin{tabular}{lr}
        \Xhline{2\arrayrulewidth}
        Phoneme embedding / FFT input block size    & 256 \\
        FFT block Conv1D filter size                & 1024 \\
        FFT Conv1D kernel size                      & 9 \\
        Number of attention heads                   & 2 \\
        Number of FFT blocks in an encoder          & 4 \\
        FFT block dropout                           & 0.1 \\
        Batch size (long context)                   & 60 (45) \\
        Batch size, duration model (long context)   & 48 (24) \\
        Optimizer, on defaults                      & Adam \\
        Learning rate                               & 2e-05 \\
        Learning rate, duration model               & 1e-05 \\
        Acoustic model checkpoint (long context)    & 200k (120k) \\ \hline
    \end{tabular}}
    \caption{Hyperparameters and model selection.}
    \label{tab:hyperparams}
\end{table}

During synthesis, we use a similar concatenation scheme.
We highlight the importance of using a similar concatenation scheme during both training and synthesis, as this results in comparable data distributions. We examine this in more detail in \S\ref{subsec:eval_infonly}.

\section{Evaluations}
\label{sec:evaluations}

We conducted experiments on internal datasets of three female English speakers referred hereon as speakers A, B, and C. By design, the data contain extensive contiguous chunks of text.
We sampled the audio at 24~kHz and extracted 80-band mel-spectrograms with a frame shift of 12.5~ms. The training sets of speakers A, B, and C consist of recordings of multi-sentence chunks taken from books with total durations of about 24~h, 32~h, and 22~h, respectively.

Table~\ref{tab:hyperparams} shows our hyperparameter and model selection choices. We train duration models for at most 2M steps and select the checkpoints with the lowest validation set mean absolute error.

\subsection{Subjective evaluations with MUSHRA tests}

\begin{table}[]
    \captionsetup{font=footnotesize}
    \centering
    \begin{adjustwidth}{-7mm}{}
    \resizebox{.47\textwidth}{!}{
    \begin{tabular}{llrrr}
    \Xhline{2\arrayrulewidth}
                    &   & \multicolumn{3}{c}{mean MUSHRA scores} \\
                    &   & this & +BERT  & reference \\ \hline           %
    Baseline        &   & $71.6\pm0.7$ & $\bf74.1\pm0.7$ & $75.9\pm0.7$ \\ %
    + multi-speaker & \footnotesize{\textsc{mt}} & $67.5\pm0.9$ & $\bf72.8\pm0.8$ & $75.3\pm0.8$ \\ \hline  %
    \end{tabular}}
    \vspace{1mm}
    \caption{Ablation results for BERT word embeddings.
    this=the system in the leftmost column, +BERT=the system in the leftmost column extended with BERT word embeddings, reference=the reference mel-spectrograms vocoded with a parallel student vocoder.
    In each row, all means are statistically significantly different under a paired two-sided t-test with $\alpha=0.01$.
    }
    \label{tab:bert_ablations}
    \end{adjustwidth}
\end{table}

\begin{table}[]
    \captionsetup{font=footnotesize}
    \centering
    \begin{adjustwidth}{-7mm}{}
    \resizebox{0.51\textwidth}{!}{
    \begin{tabular}{llrrr}
        \Xhline{2\arrayrulewidth}
        & &      \multicolumn{3}{c}{mean MUSHRA scores} \\
        & & this & \textsc{mltb} & reference \\ \hline %
        Baseline & %
        & $70.0\pm0.8$ & $\bf73.6\pm0.8$ & $75.1\pm0.8$ \\ %
        + multi-speaker & \footnotesize{\textsc{mt}} %
        & $66.9\pm0.8$ & $\bf72.9\pm0.7$ & $73.7\pm0.7$ \\ %
        + multi-speaker, BERT & \footnotesize{\textsc{mtb}} %
        & $72.1\pm0.8$ & $\bf73.4\pm0.7$ & $74.3\pm0.7$ \\ \Xhline{.5\arrayrulewidth} %
        CC2 & & $69.2\pm0.8$ & $\bf 71.0\pm0.8$ & $72.3\pm0.8$ \\ \hline
    \end{tabular}}
    \vspace{1mm}
    \caption{Results for multi-speaker long-context Transformer-BERT (MLTB).
    In each row, all means are statistically significantly different under a paired two-sided t-test with $\alpha=0.01$.}
    \label{tab:ablations}
    \end{adjustwidth}
\end{table}

We conduct MUSHRA evaluations with 24 testers. We evaluate on two voices (A and B), with 25 test samples per voice. The samples are recordings of contiguous multi-sentence excerpts from books with an average duration of approximately 19 seconds. We ask the testers to rate the naturalness of the voices reading the samples.
We vocode with a parallel student vocoder \cite{Jiaoetal2021}.
To nullify the effect from the vocoder, we also apply it to the mel-spectrograms extracted from the reference audio. We use these samples as the upper anchor in the MUSHRAs, and we refer to them as ``reference'' in the tables above.

\textbf{BERT word embeddings.} Adding BERT word embeddings leads to more natural speech (cf.\ strong statistical gains in mean MUSHRA scores in Table~\ref{tab:bert_ablations}).
While this agrees with the literature on BERT for single-speaker models, this also appears to be the case for multi-speaker systems (the bottom row of Table~\ref{tab:bert_ablations}, multi-speaker Transformer (MT) vs multi-speaker Transformer-BERT (MTB)). Interestingly, the linguistic information available in BERT cannot be compensated by an almost threefold increase in training data due to multiple speakers.
We note, though, that the MT samples had some harshness (not reported for the other systems), which likely impacted the scores for this system.

\textbf{Ablations on multi-speaker long-context Transformer-BERT.} Next, we compare various model combinations to our main system with all three extensions, multi-speaker long-context Transformer-BERT (MLTB). We observe consistent statistically significant improvements in all ablations (Table~\ref{tab:ablations}). They are particularly large due to the length of the test samples, which comprise about 4 sentences on average and, thus, favor long-context systems.
We highlight the fact that the addition of long context training and synthesis leads to a 59.0\% gap reduction over the MTB system, featuring BERT word embeddings. Looking at per-voice results, this gain is much larger for voice B than voice A, which could be due to the textual differences in the data.

\textbf{Comparison to a strong multi-speaker system.} We also evaluate MLTB against CC2 (the bottom row of Table~\ref{tab:ablations}), a highly competitive autoencoder-based single-utterance multi-speaker system \cite{Karlapatietal2022}. CC2 learns a word-level variational autoencoder to capture word prosody. At synthesis time, a separate model predicts autoencoder bottleneck features from fine-tuned BERT word embeddings. We observe a strong statistically significant improvement from MLTB (a gap reduction of 58.3\%).

\subsection{Further analysis}

\subsubsection{Impact on phoneme durations and pausing}

In this section, we investigate the impact of the proposed extensions on phoneme durations predicted by the duration model.
We examine how well predicted durations match the reference test-set distribution, looking separately at non-pause phonemes, inter- and intra-sentence pauses.
The correct placement and durations of pauses are particularly important for the overall intelligibility and coherence of speech \cite{Klimkovetal2017}.
We quantify duration error using mean squared error (MSE) and the coefficient of determination ($R^2$).

\begin{table}[]
\captionsetup{font=footnotesize}
\centering
\resizebox{.49\textwidth}{!}{
\begin{tabular}{llr?rrr}
\Xhline{2\arrayrulewidth}
\multicolumn{1}{l}{\multirow{3}{*}{}} &
\multicolumn{2}{r?}{non-pauses} &
\multicolumn{3}{c}{pauses}  \\
& & & \multicolumn{1}{r}{within} & \multicolumn{2}{c}{between}           \\
& \multicolumn{1}{c}{}                        & \multicolumn{1}{r?}{MSE}        & \multicolumn{1}{r}{MSE}    & \multicolumn{1}{r}{MSE} & \multicolumn{1}{r}{$R^2$} \\ \hline
Baseline    &                          & 8.8 & 57.3 & 1030.1 & -0.19 \\
+ multi-speaker & \footnotesize{\textsc{mt}}                       & 8.3 & 50.4 & 923.0  & -0.07 \\
+ multi-speaker, BERT & \footnotesize{\textsc{mtb}}                & 8.3 & 47.3 & 755.9  & 0.12  \\
+ multi-speaker, BERT,    & \multirow{2}{*}{\footnotesize{\textsc{mltb}}}\quad\quad           & \multirow{2}{*}{\bf 7.2} & \multirow{2}{*}{\bf 44.3} & \multirow{2}{*}{\bf 680.8} & \multirow{2}{*}{\bf 0.21} \\
\quad\quad long context               &     &      &   \\ \hline
\end{tabular}
}
\vspace{1mm}
\caption{Comparison of objective metrics between predicted and reference durations for various model configurations. within=intra-sentence pauses, between=inter-sentence pauses, 
MSE=mean squared error, $R^2$=coefficient of determination.}
\label{table:objective_metrics}
\end{table}

Generally, we see progressive improvement in durations of both pauses and non-pauses as the extensions get incorporated progressively into the baseline (Table~\ref{table:objective_metrics}).
We highlight a large improvement in pause durations from a multi-speaker model over the single-speaker baseline.

\begin{figure}
    \captionsetup{font=footnotesize}
    \centering
    \includegraphics[scale=.35]{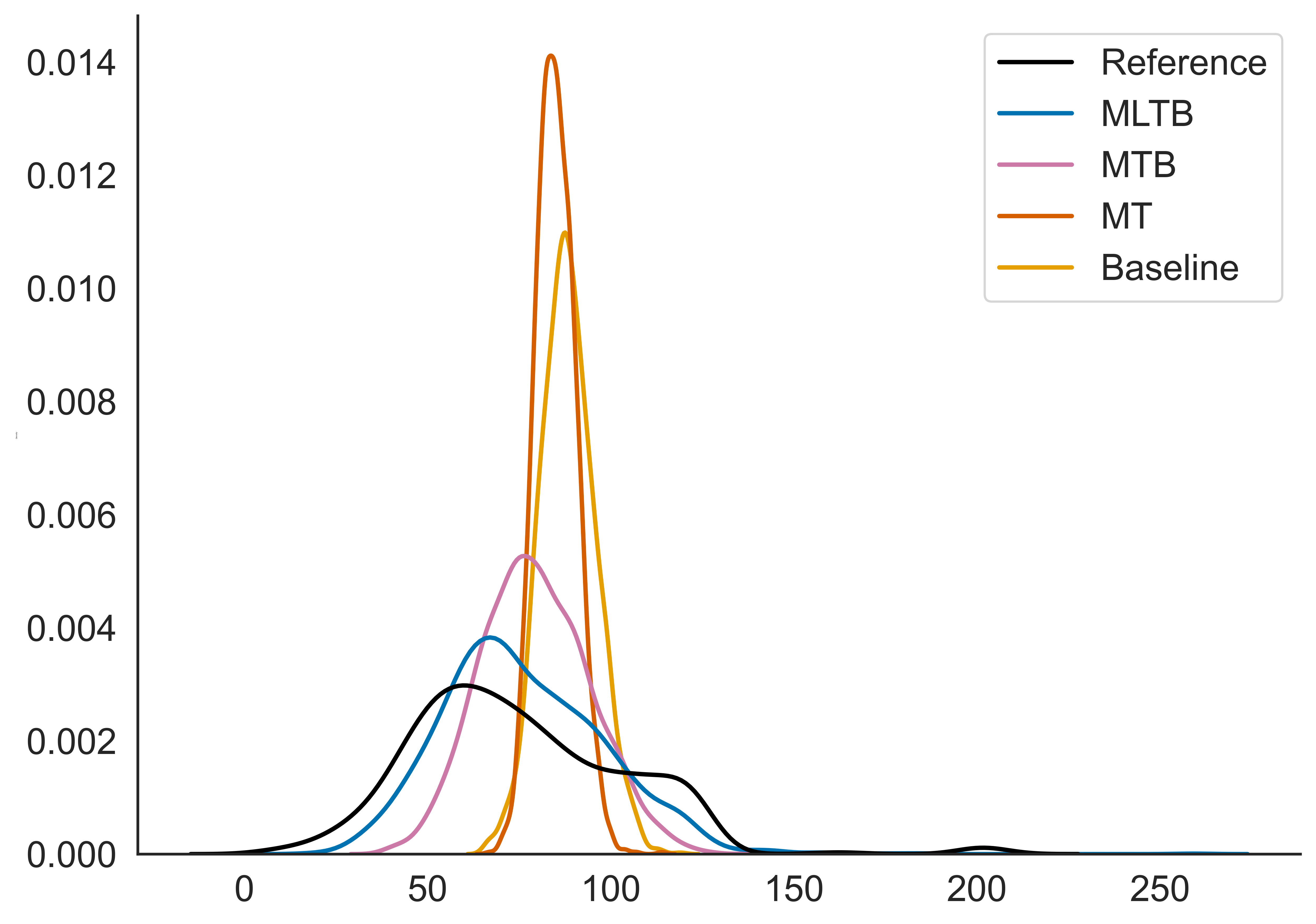}
    \caption{Distribution of inter-sentence pauses in frames.}
    \label{fig:intersentences}
\end{figure}

We also note the large absolute improvement in pauses between sentences from the addition of word embeddings.
The negative $R^2$ values for models without word embeddings or long context indicate they fit inter-sentence pauses worse than the test set mean. 
Long-context models are better at explaining variance of inter-sentence pause durations. Figure~\ref{fig:intersentences} shows distributions of inter-sentence pause durations for various model combinations. The models without word embeddings (the baseline and MT)
produce dominant unimodal distributions. A significant improvement comes from long context as these models come closest to the reference distribution.

\subsubsection{Manual error inspection}

We examine test samples with large MUSHRA score gaps between any two systems. We highlight the following case where long context leads to strong gains:
The sample is a sequence of many shorter sentences; or the sample contains a discontinuity which requires special prosody (e.g. a topic change). %

We also look into MUSHRA score differences attributable to BERT word embeddings. BERT-powered systems have fewer problems with general vs wh-questions, and do better on phonetically ambiguous input (e.g.\ the sentence-final ``too'') and sentences with complex syntax,
improving overall comprehensibility with appropriate pausing and a slower pace.

\subsubsection{Effect of longer context during training and synthesis}
\label{subsec:eval_infonly}

We also examine whether a system trained on short, single-sentence input can be directly used to synthesize multi-sentence input.
To this end, we conducted a preference test between MLTB and MTB with which we ran synthesis on chunks of concatenated sentences (MTB-synL). The preference test consisted of 50 utterances, 25 each from speakers A and B, of approximately 19 seconds each on average, which were rated by 120 crowdsourcing platform testers. We found that MLTB performs statistically significantly better than MTB-synL ($\alpha = 0.05$).
Upon further inspection, we found that some of the samples from MTB-synL were suffering from glitches around sentence boundaries. We attribute this to the mismatch in the distributions of inputs to the MLTB-synL system at training and synthesis.

\section{Conclusion}

This paper looks at TTS for domains with multi-sentence input.
To increase the coherence and contextual appropriateness of prosody %
on this type of input, we examine three simple and effective extensions to a Transformer-based FastSpeech-like baseline: long context, contextual word embeddings, and multi-speaker modeling. We show that all these extensions contribute to large improvements in objective metrics relating to pausing and phoneme durations, as well as in subjective evaluations for speech naturalness. We report strong synergies. Our best system, which combines all three extensions, leads to statistically significant improvements in speech naturalness over all its competitors.

\section{Acknowledgement}
We would like to thank Simon Slangen, Bajibabu Bollepalli, Arent van Korlaar, and Ray Li for their help with this work. 

\bibliographystyle{IEEEtran}

\bibliography{refs}

\end{document}